\documentclass{emulateapj}
\usepackage{natbib}
\shorttitle{Magnetic turbulence from streaming cosmic rays}
\shortauthors{Stroman et al.}

\newcommand{\ts}{}

\begin{document}

\title{Kinetic simulations of turbulent magnetic-field
growth by streaming cosmic rays}

\author{Thomas Stroman and Martin Pohl\altaffilmark{1}}
\affil{Department of Physics and Astronomy, Iowa State University,
    Ames, IA 50011}

\and

\author{Jacek Niemiec}
\affil{Institute of Nuclear Physics PAN, ul. Radzikowskiego 152, 31-342 Krak\'{o}w, Poland}
\altaffiltext{1}{Now at: Institut f\"ur Physik und Astronomie, Universit\"at
Potsdam, 14476 Potsdam-Golm, Germany;\\
and DESY, 15738 Zeuthen, Germany}

\begin{abstract} % needs work, more detail
Efficient acceleration of cosmic rays
(via the mechanism of diffusive shock acceleration)
requires turbulent, amplified magnetic fields
in the shock's upstream region. We present results of multidimensional
particle-in-cell simulations aimed at observing the magnetic field
amplification that is expected to arise from the cosmic-ray current
ahead of the shock, and the impact on the properties of the upstream
interstellar medium. We find that the initial structure and peak strength
of the amplified field is somewhat sensitive to the choice of parameters,
but that the field growth saturates in a similar manner in all cases:
the back-reaction on the cosmic rays leads to modification of their
rest-frame distribution and also a net transfer of momentum
to the interstellar medium, substantially weakening their relative drift
while also implying the development of a modified shock. The upstream
medium becomes turbulent, with significant spatial fluctuations in
density and velocity, the latter in particular leading to
moderate upstream heating; such fluctuations will also have
a strong influence on the shock structure.
\end{abstract}

\section{Introduction}
The forward shocks of young shell-type supernova remnants may
be efficient acceleration sites for Galactic cosmic rays. The theory of diffusive
shock acceleration (DSA) provides a promising mechanism by which a small fraction
of particles can attain nonthermal energies in such an environment,
provided they are confined to the shock vicinity; the review by \citet{2008ARA&A..46...89R}
and its references have treated the subject in detail.
This confinement may be supplied to some degree by the interstellar magnetic field,
but the estimated upper limit in energy falls short of what is observed in cosmic rays
of Galactic origin. The implied need for a much stronger, turbulent upstream field, together with
observational evidence for strong magnetic
fields immediately behind the forward shock, has attracted attention
to the question of upstream magnetic-field amplification.

It is well known that ion beams can generate magnetic turbulence via
resonant and non-resonant interactions \citep{1984JGR....89.2673W}. \citet{2004MNRAS.353..550B} (hereafter B04)
suggested that non-resonant magnetic field amplification may also be
caused by the cosmic-ray current expected in the cosmic-ray
precursor of a quasi-parallel shock.
Although substantial field amplification has occurred in
magnetohydrodynamical (MHD) simulations that assume a constant
cosmic-ray current \citep[B04; also][]{2008ApJ...678..255Z},
the approximations of MHD may not
be appropriate for modeling the non-linear evolution and eventual saturation
of this instability; we must use kinetic methods to simulate these later stages
with accuracy, including any back-reaction on the cosmic rays.

Our earlier work \citep[][hereafter N08]{2008ApJ...684.1174N} used 2-D and 3-D
particle-in-cell (PIC) simulations to model the growth and saturation
of a current-driven instability.
We relaxed the condition assumed in the calculations of B04 that
the rate of unstable growth must be much less than the ion gyrofrequency,
and we found that an oblique filamentary mode dominated the initial
magnetic field growth. \citet{2009ApJ...694..626R} confirmed the findings
of N08, and in particular investigated the parameters for the
transition between the filamentation mode seen in N08 and the
parallel-wave mode seen in the MHD simulations,
and verified that the parallel mode appears only in the
regime with $\omega \ll \Omega_i$.
However, despite the difference in initial unstable modes,
the non-linear characteristics of the system observed in both works
\citep[N08;][]{2009ApJ...694..626R} were similar.
In particular, both investigations observed only modest amplification
of the magnetic field, and saturation by the same mechanism.  The
analytical calculations of \citet{2009MNRAS.397.1402L}, with a
kinetic treatment of the current-driven instability, predict
a saturation level consistent with the kinetic simulations.

In this paper, we report more recent work, in which we simulate the current-driven instability and its
saturation in a parameter regime in which the non-relativistic
ion gyrofrequency far exceeds the (complex) frequency of the
current-driven instability. Such parameters are particularly relevant
to the environment of non-relativistic strong shocks whose cosmic-ray precursors
are sparse in comparison with the magnetized ambient interstellar medium (ISM);
an understanding of the limits of field amplification in such an environment
is therefore essential to the formulation of a self-consistent theory
of cosmic-ray acceeleration in the Galaxy. {\ts We find that the initial
evolution of the magnetic field in this parameter regime is in agreement
with the predictions of B04 and consistent with the transition parameter, $\omega/\Omega_i$,
identified by \citet{2009ApJ...694..626R}. We observe that the field growth
is limited to modest values by the same saturation mechanism as before
\citep[N08;][]{2009ApJ...694..626R}.
In addition to the magnetic field, we examine the evolution of the
upstream medium-precursor system in terms of
bulk and local properties that are relevant for modeling the shock acceleration process.
Notable findings include the rise of strong turbulent motion
relative to the bulk upstream rest frame, and of motional electric fields
associated with the magnetic fields that permeate the
turbulent flow. The temperature of the upstream medium, measured
in the local rest frame as determined with high spatial resolution, attains
significant values during the evolutionary phase characterized by the turbulent
motion, such that the ratio of magnetic to thermal energy density
may in fact decrease as a result of this precursor-ISM interaction. Such
modifications to the upstream medium may lead
to substantial effects on the propagation of the subshock, and on the properties of
the downstream medium: both the pre-existing upstream turbulence
after advection and shock compression, and further magnetic-field
generation via turbulent dynamo processes, may be quite sensitive
to turbulence in the upstream medium
\citep{2007ApJ...663L..41G,2009ApJ...695..825I}.}

We begin by describing in \S \ref{model} the simple physical scenario
that captures all the ingredients necessary for the streaming instability
we want to simulate,
with \S \ref{implementation} summarizing our computational representation
of that environment,
and the specific parameterizations being outlined in \S \ref{descriptions}.
We present our simulation results in \S \ref{result} along with discussion
of their significance, and finally summarize in \S \ref{conclusions}.

\section{\label{setup}Simulation setup}
\subsection{\label{model}Model}
To study the growth and saturation of magnetic turbulence in response
to a cosmic-ray current, we simulate the evolution
of a system representative of the upstream environment of a
non-relativistic strong parallel shock.
We assume that the interstellar medium is a
homogeneous collisionless plasma composed entirely of ions and electrons
(number density $N_i$ and $N_e$, respectively)
of opposite but equal charge $\pm e$,
permeated with a uniform magnetic field $B_{\parallel 0}$.
A population of cosmic ray ions (density $N_{CR}=N_e-N_i$)
drifts at the shock velocity $v_{sh}$ along the magnetic field
relative to the plasma
and carries a current density $j_{CR}=eN_{CR}v_{sh}$;
the slight electron excess in the plasma maintains charge neutrality,
and the electrons drift relative to the ions with $v_d=v_{sh}N_{CR}/N_e$
to provide a return current density $j_{ret}=-e N_e v_d=-j_{CR}$.
The ISM ion and electron populations are each
distributed according to Maxwell-Boltzmann statistics in their
respective rest frames with the same temperature,
given by the electron thermal velocity $v_{e,th}$.

Two assumptions in our setup,
that only electrons provide the return current and
that the streaming cosmic rays consist only of ions,
arise from the large mass ratio $m_i/m_e$
between physical ions and electrons.
The electrons will respond to any local charge or current imbalance
on a much shorter timescale than the ions will.
Moreover, at high energies, ions outnumber electrons (on account of
their greater rigidity) by a factor that depends on the spectral
index of their momentum distribution but is generally large enough that
 cosmic-ray electrons play a negligible role in the precursor physics.
Cosmic rays upstream undergo pitch-angle scattering
off magnetic inhomogeneities and
approach a quasi-isotropic distribution,
forming a steady cosmic-ray precursor in the shock rest frame.
Ions with lower energies will be advected back toward the shock
sooner than ions with higher energies,
so at some distance upstream one expects to find a population of
high-energy cosmic-ray ions drifting with the shock,
up to some cutoff energy and/or distance \citep{2009ApJ...694..951R}
associated with particles that escape the system.

For our simulation, we simplify the cosmic-ray distribution function.
We assume that the isotropic, scattered cosmic rays of the precursor
are energetically much more significant than any escaping cosmic rays
that happen to be passing through the region of our simulation,
so we neglect the latter altogether.
Our box size is assumed to be small in comparison both to
the scale length of the cosmic-ray precursor,
such that we may neglect spatial variations in their distribution, and
to the Larmor radii of any cosmic rays present,
such that the details of their energy distribution are similarly negligible,
and we replace the already-narrow range of energies with
a single energy (in the CR population rest frame),
with Lorentz factor $\gamma_{CR}$.
The quasi-linear calculations in B04 do not depend explicitly on cosmic-ray energy,
and tests with various values of $\gamma_{CR}$ have not
produced significantly different outcomes,
so it is unlikely that our choice of energy distribution
obscures behavior that a more realistic model would illuminate.

In considering the current carried by cosmic rays through
a cold ambient plasma parallel to a uniform magnetic field,
the calculations in B04 predict an instability in which the dominant mode,
circularly polarized magnetic fluctuations with
wavenumber $k_{\parallel max}$,
grows at a rate $\Im \omega = \gamma_{max}=v_A k_{\parallel max}$,
$\Re \omega \approx 0$, where
\begin{equation}
\label{instab}
k_{\parallel max}=\frac{e N_{CR}v_{sh} B_{\parallel 0} }{2 N_i m_i v_A^2}
\approx\frac{\mu_0 J_{CR}}{2B_{\parallel 0}}
\end{equation}
and
$v_A=\left[B^2_{\parallel 0}/\mu_0\left(N_e m_e+N_i m_i\right)\right]^{1/2}$
is the Alfv\'en velocity of the plasma.
As mentioned above, this result is subject to the restriction that
the fluctuation satisfy $\omega\ll \Omega_i$;
that is, that the growth rate of the instability
must be much less than the nonrelativistic ion gyrofrequency.

\begin{table*}[th]

\begin{center}
\caption{Parameters and selected results of simulations}
\label{table:runs}
\begin{tabular}{lcccccccccc}
Run & Grid ($\Delta^2$) & $N_i$ ($\Delta^{-2}$) & $\lambda_{se}$ ($\Delta$) &  $v_{sh}/c$
& $\lambda_{max}$ ($\lambda_{se}$) & $\gamma_{CR}$ & $B_{\perp}/B_{\parallel 0}$ & $\gamma/\gamma_{max}$ & $\gamma_{max}/\Omega_i$
\\
\hline
A & $9000\times 2400$ & 25 & 4 & 0.4 & 111 & 50 & 13.3 & 0.96 & 0.4  \\
A$'$ & $3000 \times 1600$ & 25 & 4 & 0.4 & 111 & 50 & 17.6 & 0.86 & 0.4  \\
B & $4500 \times 1800$ & 20 & 6& 0.4 & 111 & 50 & 19.7 & 0.89 & 0.4  \\
C & $4000 \times 1500$ & 25 & 5 & 0.3 & 148 & 10 & 11.6 & 0.78 & 0.3  \\
C$'$ & $4000 \times 1500$ & 25 & 5 & 0.3 & 148 & 10 & 9.9 & 0.93 & 0.3 \\
D & $3200 \times 1200$ & 25 & 5 & 0.4 & 111 & 10 & 12.7 & 0.86 & 0.4  \\
E & $3200 \times 1200$ & 25 & 5 & 0.4 & 111 & 25 & 18.4 & 0.90 & 0.4  \\
F & $3200 \times 1200$ & 25 & 5 & 0.4 & 111 & 200 & 20.1 & 0.91 & 0.4  \\

\hline

\end{tabular}
\tablecomments{Run A$'$ reproduces the parameters
of Run A on a spatial scale comparable to run B. Run C$'$ differs from Run C only in
that the cosmic-ray particles are not split into multiple simulated particles. $N_i$ is the number
of ions per cell (size $\Delta^2$), $\lambda_{se}=c/\omega_{pe}$ is the electron skin depth, $v_{sh}$
is the shock speed at which the cosmic-ray population drifts; $\gamma_{CR}$ is the cosmic-ray Lorentz
factor in the population rest frame. $\gamma_{max}/\Omega_i$ is the ratio of
the predicted growth rate of the most
unstable mode (with wavelength $\lambda_{max}$) \citep{} to
the nonrelativistic ion gyrofrequency. The
peak measured perpendicular magnetic field and growth rate are given by $B_\perp$ and $\gamma$ in units
of their respective reference quantities. Parameters common to all runs include an ion-to-electron
mass ratio $m_i/m_e=50$, a density ratio of ions to cosmic rays $N_i/N_{CR}=50$, and an initial
homogeneous magnetic field antiparallel to the cosmic-ray drift
such that the Alfv\'en speed is $v_A=0.01c$.}
\end{center}
\end{table*}

\subsection{\label{implementation}Implementation}

Our simulations employ a modified version of the particle-in-cell code
TRISTAN \citep{1993cspp.book......M}.
The code is updated to work in a high-performance computing environment
and uses the charge-conserving current deposition algorithm of
\citet{2003CoPhC.156...73U}.
We have incorporated a fourth-order-accurate algorithm for updating
the electromagnetic fields based on Maxwell's equations \citep{2004JCoPh.201..665G},
and adapted the code to a truly 2.5-dimensional simulation.\footnote{As reported
in N08, we have previously approximated two dimensions by simulating
a volume with a thickness of only three grid cells.}

The shortest relevant physical time- and length-scales are given by the
electron plasma period $\omega_{pe}^{-1}=\left[m_e \epsilon_0 /e^2 N_e\right]^{1/2}$
and electron skin depth $\lambda_{se}=c/\omega_{pe}$, respectively.
To ensure that physical small-scale electrostatic effects
are distinct from those arising from the
artificial discretization of space and time,
for our simulations we choose
parameters such that $\lambda_{se}\ge4\Delta$,
where $\Delta$ is the length of one cell,
and $\omega_{pe}^{-1}\ge 8.9\delta t$,
where $\delta t$ is the timestep size.
The wavelength $\lambda_{max}\equiv 2\pi/k_{\parallel max}$
of the predicted magnetic-field fluctuations is of order $10^2 \lambda_{se}$
for all simulations (see Table \ref{table:runs}).
We perform our simulations in a rectangular computational grid,
with both the homogeneous magnetic field and the
direction of cosmic-ray drift aligned with the longer dimension.
Periodic boundary conditions in both directions allow particles and fields
to ``wrap around'' to the far side of the grid,
but this also imposes an upper limit to the size of structures
that can be represented accurately.
Having observed a migration of the dominant structures toward
increasing size scales in N08, for one simulation (Run A) we
chose a grid slightly larger than $20 \lambda_{max}$
in the drift direction,
and $5 \lambda_{max}$ in the transverse direction.
We ran additional simulations on smaller grids to explore the
influence of variations in our choice of parameters on the initial
amplification of the magnetic field and the turbulent properties
of the field and plasma.

 \begin{figure}[th]
  \centering
  \includegraphics[width=3in]{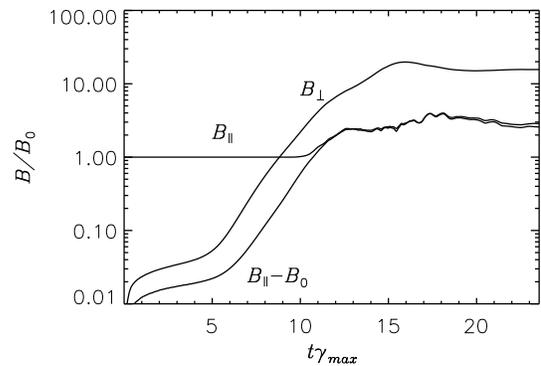}
  \caption{Evolution of the magnetic field components in Run B displaying the growth and saturation
of the instability predicted in B04.
The magnetic field amplitude
is in units of the initial homogeneous field, and time is in dimensionless units set
by the calculated inverse growth rate $\gamma_{max}^{-1}$ from that instability.}
  \label{famp}
 \end{figure}

To minimize statistical noise associated with the particles,
we employ a combination of large per-cell particle counts
and low-pass filtering of the electric currents arising from their motion.
We initialize the simulations with
either 20 or 25 plasma ions per cell, and a density ratio of
$N_i/N_{CR}=50$ plasma ions per
cosmic-ray ion. However, in most runs we split each cosmic ray
into ten simulated particles that preserve the charge-to-mass ratio
but provide better statistics, and the contribution
of each particle to currents and densities is weighted accordingly.

Electrons, less massive than the ions by a factor of $m_i/m_e=50$ in all tests,
are physically one population but simulated as two populations.
In a simulation for which $N_i=25$, each cell contains 25 ``full'' electrons
as well as
five simulated particles, which add up to half of a split electron.
The two simulated electron populations are initialized
according to the same distribution function, and
behavioral comparisons of the full electrons and the
split electrons in the simulations reveal no differences
between the two populations, to within statistical limits.

The initial temperature of the plasma is set artifically high, with $v_{e,th}=0.01c$, in
order to mitigate Buneman-type electrostatic effects arising from the drift between the
stationary population of ions and the slowly drifting electrons. Our simulations in N08
demonstrated that cooler plasmas were heated through such effects on a much shorter
timescale than any turbulent magnetic-field amplification, but the resulting anisotropy
of the ion distribution function in particular persisted on such timescales. A higher
initial temperature leads to better preservation of isotropy against the intra-plasma
drift. Additionally, using a density ratio $N_i/N_{CR}$ of 50 instead of 3 (the
value in N08) significantly reduces $v_d$.

The cosmic rays are initialized to be mono-energetic
and isotropic in the shock rest frame. This frame drifts in the $-x$-direction,
antiparallel to the homogeneous magnetic field, at a speed $v_{sh}=0.4c$ or $0.3c$.
We set the magnetic field strength such that the Alfv\'en speed $v_A=0.01c$,
or equivalently that the plasma frequency exceeds the
Larmor frequency by a factor of $\omega_{pe}/\Omega_e\approx 14.14$.

\subsection{\label{descriptions}Run-specific parameters}
In Table \ref{table:runs} we summarize the initial conditions and
some observations from each of the simulations we discuss. The
default parameters are described below, followed by a list of
each secondary run's purpose and any deviations from those parameters.

Run A is our large-grid run for this paper, and is the basis for
many subsequent ``test'' simulations.
The cosmic rays are isotropic and mono-energetic with Lorentz factor
$\gamma_{CR}=50$ in the CR rest frame,
which drifts at $v_{sh}=0.4c$ in the $-x$-direction in the simulation frame.
There are initially 25 plasma ions per cell, and each cosmic-ray ion
is split into ten equal parts, for a net of five simulated particles per cell.
For these parameters, $\lambda_{max}=444\Delta$.
In order to capture the non-linear stage when the dominant structure
size exceeds this, we simulate on a grid of $9000\Delta \times 2400 \Delta$,
corresponding to $20.26 \lambda_{max}\times 5.4\lambda_{max}$.

Run A$'$ uses the same physical parameters as Run A,
but on a smaller grid,
fitting $6.75\lambda_{max} \times 3.6\lambda_{max}$.

Run A uses a large grid to ameliorate the limitations imposed
by periodic boundary conditions, which is particularly relevant
during the later evolution when structures in the magnetic field
begin to grow quite large. However, for the purpose of analyzing the properties
of the upstream medium during the growth and saturation, we simulate a smaller
region with
a greater detail level.
Run B is our test with increased spatial resolution, in which the
electron skin length has been increased to $\lambda_{se}=6\Delta$,
and the particle density reduced to $N_i=20$,
while holding the other parameters fixed. The grid is
$6.75\lambda_{max} \times 2.7\lambda_{max}$.

Runs C through F are tests with varying cosmic-ray Lorentz factors.
All use the intermediate value for electron skin length of
$\lambda_{se}=5\Delta$.

Run C has a reduced shock speed $v_{sh}=0.3c$.
Its cosmic rays also have a reduced Lorentz factor, $\gamma_{CR}=10$.
It is unique among our runs in that it does not split
the cosmic rays. The grid for Run C admits
$5.4\lambda_{max} \times 2.03\lambda_{max}$.
Run C$'$ is identical to Run C but splits the cosmic rays.

Runs D, E, and F use the usual shock speed $v_{sh}=0.4c$
and cosmic-ray Lorentz factor $\gamma_{CR}=10$, 25, and 200,
respectively. Their grid measures
$5.76\lambda_{max} \times 2.16\lambda_{max}$.

 \begin{figure*}[th]
  \centering
  \includegraphics[width=6in]{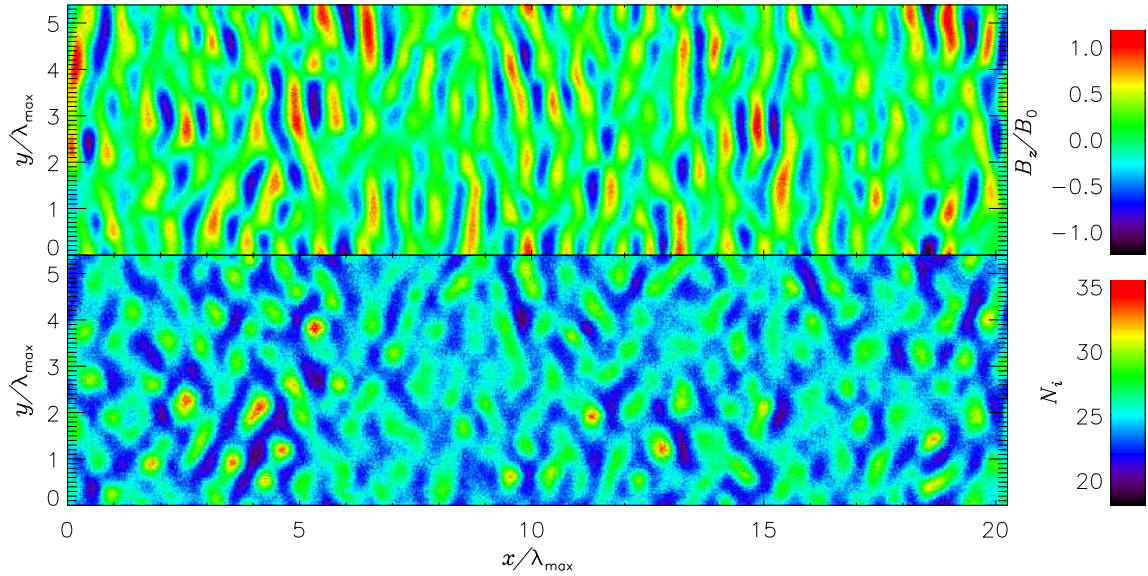}
  \caption{The perpendicular magnetic field in units of $B_{\parallel 0}$ (top) and density of plasma ions (bottom) at $t\gamma_{max}=7.0$ for Run A,
when $\langle \delta B / B_{\parallel 0}\rangle \approx 0.50$.
The parallel-mode structure is clearly visible in the magnetic field. Localized
density fluctuations about the mean (25 ions per cell) are smaller, with
an amplitude of $\langle \delta N_i / N_i \rangle \approx 0.08$.}
  \label{bzni_lin}
 \end{figure*}

 \begin{figure*}[th]
  \centering
  \includegraphics[width=6in]{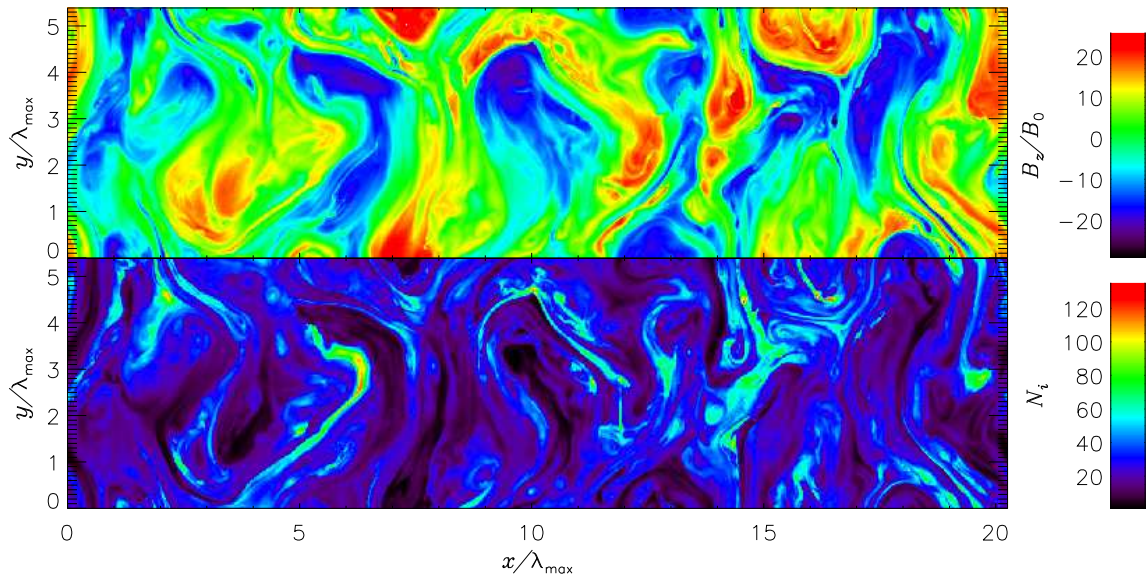}
  \caption{As in Figure \ref{bzni_lin}, the perpendicular magnetic field and density of plasma at $t\gamma_{max}=14.0$ in Run A.
The plasma density fluctuations are strong and well correlated with the magnetic field structures during the non-linear
stage of magnetic field growth. The structures move rapidly on oblique trajectories at this time in
the simulation (see Figure \ref{drift}).
    }
  \label{bzni_non}
 \end{figure*}
\section{Results and discussion}
\label{result}
The spatially-averaged amplitude of the turbulent components
of the magnetic field,
parallel and perpendicular to the initial
homogeneous field $B_{\parallel 0}$,
are plotted as a function of time in Figure \ref{famp} for Run B,
representing well the evolution in all runs.
The unit of choice for time is the inverse predicted growth rate
$\gamma_{max}^{-1}$. When $t\gamma_{max}\approx 5$, the
field enters a period of exponential growth.
As seen in Table \ref{table:runs},
the peak growth rate observed is often 10\%--20\% less than
$\gamma_{max}$; this small discrepancy may be primarily a thermal effect.
As discussed in N08, the finite temperature
of the plasma tends to reduce the maximum growth rate from
its value in the cold-plasma limit.

Initially, the growth of the magnetic field is stationary, in the form
of non-propagating waves with wave-vectors oriented parallel to the
cosmic-ray drift and similar in magnitude to
$k_{\parallel max}$ (Figure \ref{bzni_lin}.
The cosmic-ray Larmor radius $r_{CRg}\gg\lambda_{max}$,
so this is not the resonant interaction of cosmic rays
with self-generated Alfv\'en waves,
but is instead the non-resonant
generation of an approximately purely growing wave mode.
However, density fluctuations begin to appear while
$\delta B \ll B_{\parallel 0}$. The root-mean-square magnitude of the
normalized density fluctuations
$\langle \delta N_i / N_i\rangle$
during the linear growth stage is smaller than
$\langle \delta B / B_{\parallel 0} \rangle$,  typically at a level
of 20 to 50 \% between $t\gamma_{max}=5$ and $t\gamma_{max}=10$.
As the turbulent magnetic field becomes comparable in strength to the
homogeneous field, the plasma begins to move and the cosmic
rays slow down, reducing the current and saturating the
magnetic-field amplification at a level of 10--20 $B_{\parallel 0}$.

Any density and velocity fluctuations that arise may contribute to
further deviation from a purely growing parallel mode,
and may change the environment enough that further field amplification
is not favored. We observe that when $B_\perp\sim B_{\parallel 0}$,
the growth rate is only gently decreasing and the field still has
approximately a parallel-mode structure,
but when $B_\perp \sim 10 B_{\parallel 0}$,
the spatial organization of the field no longer resembles a parallel mode.
As seen in Figure \ref{bzni_non},
the dominant length scale of the fluctuations grows and significantly exceeds
$\lambda_{max}$. Substantial variation appears along the direction
transverse to the drift, reaching the maximum length scale allowed
by the periodic boundary conditions of even our largest computational
grid. The field growth comes to a stop, saturating generally
between 10 and 20 times the amplitude of the initial homogeneous field.

 \begin{figure*}[ht]
   \centerline{\includegraphics[width=3in]{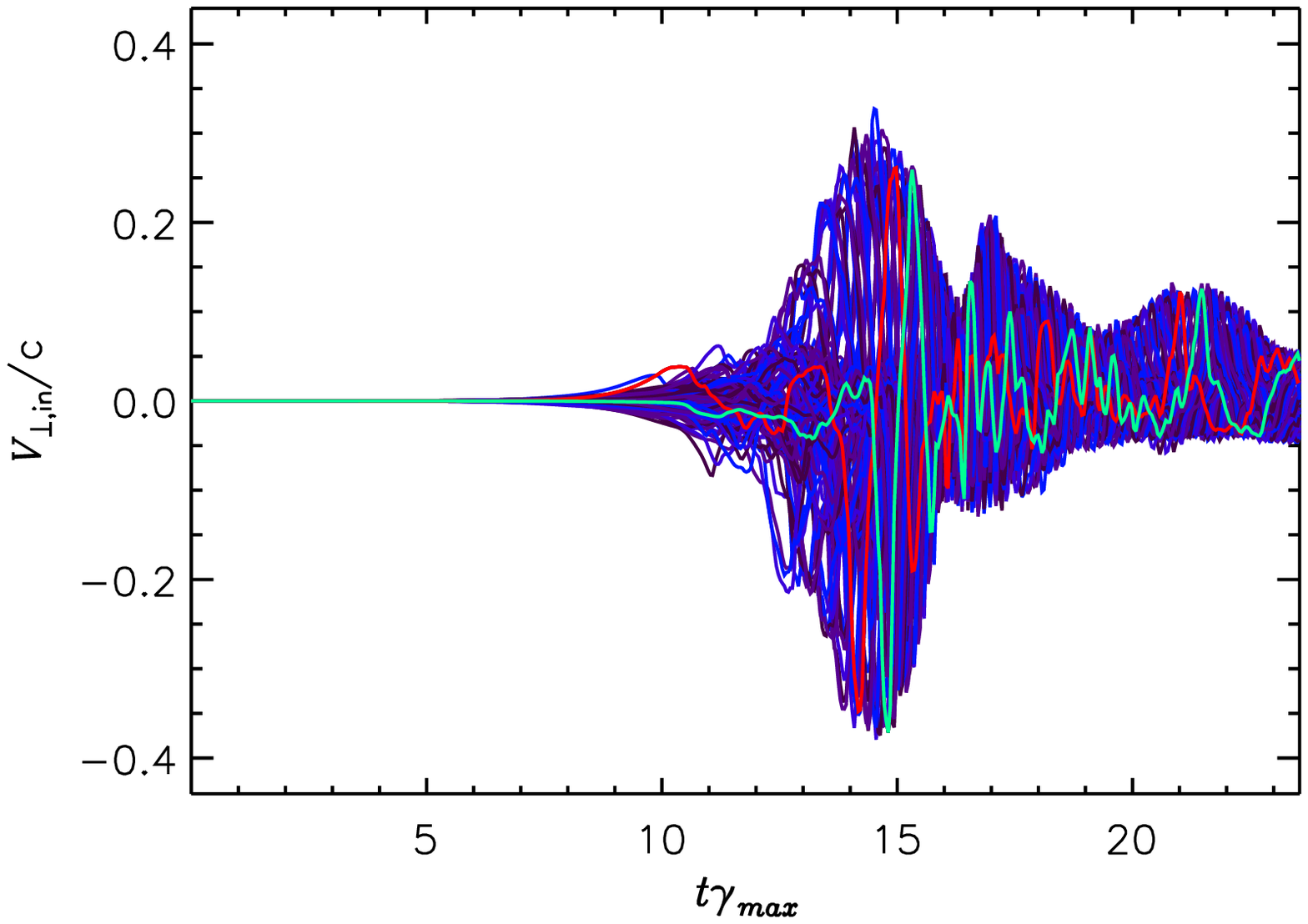}
              \hfil
              \includegraphics[width=3in]{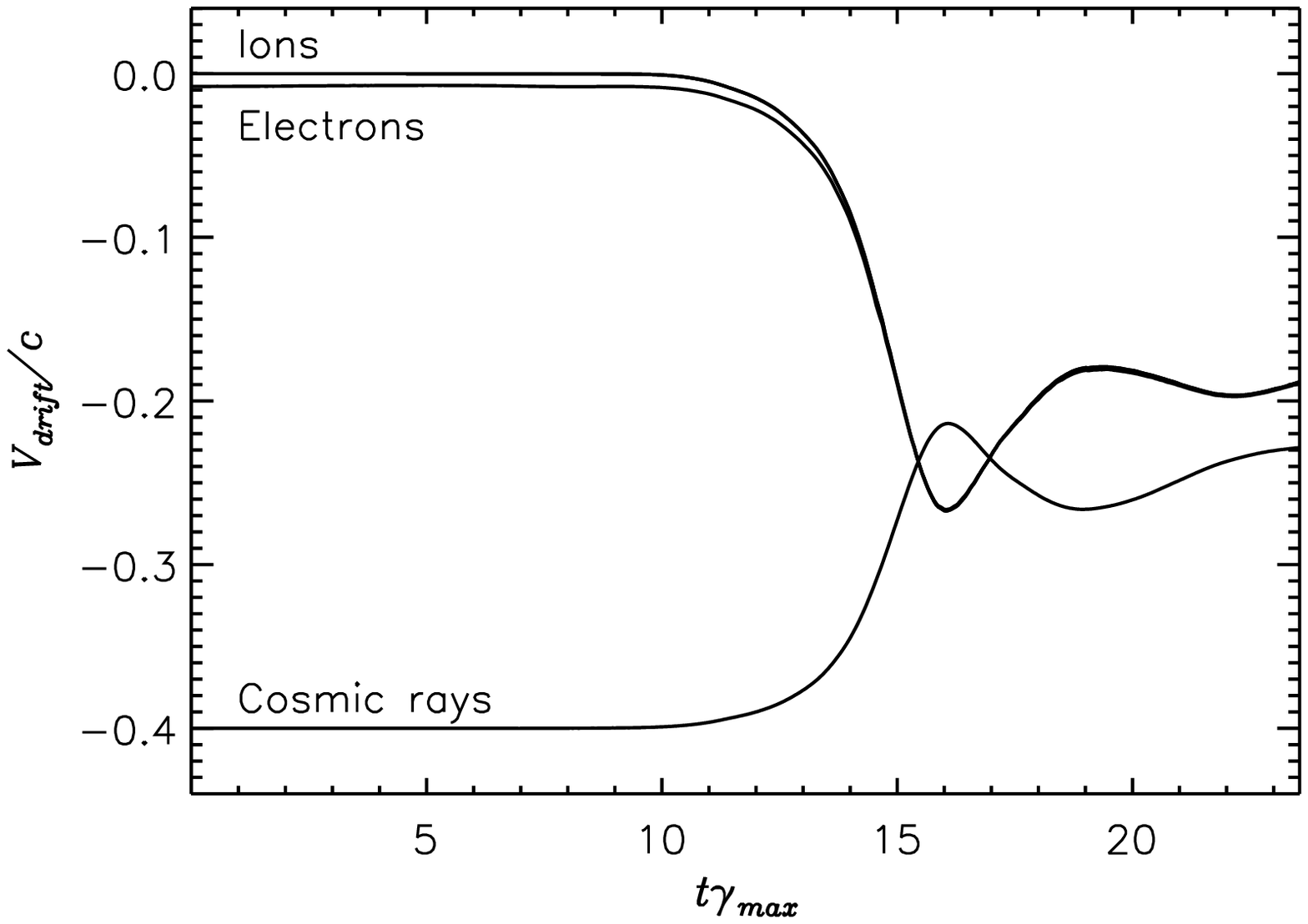}
             }
   \caption{{\bf Left:} From Run B, the transverse in-plane velocity of the plasma in regions of size
$(0.45 \lambda_{max})^2$. Two regions have been selected and superimposed in lighter colors
for clarity. {\bf Right:} In the same run, momentum transfer from the cosmic rays to the background plasma removes
their relative drift, limiting the amplification of the magnetic field. The ``overshoot''
at $t\gamma_{max}\sim 16$ appears in the small-grid runs but is not present in Run A, with the larger grid, so it is likely a consequence of the periodic boundary conditions}
   \label{drift}
 \end{figure*}

The reduction of the field growth rate accompanies the onset of
turbulent plasma motion. Parcels of ambient plasma whose size
is comparable to $\lambda_{max}$ move and collide, giving rise
to considerable density fluctuations. Figure \ref{drift} (left) indicates
that the plasma parcels can attain a considerable range of speeds
in the directions transverse to the cosmic-ray drift.
The strong perpendicular field and turbulent plasma allow bulk momentum
transfer from the cosmic rays to the ambient plasma, so superimposed on
the turbulent motions, the plasma begins to drift more and more rapidly
in the direction of the cosmic rays, while the cosmic rays simultaneously
slow down (Figure \ref{drift}, right), consistent with the quasi-linear
predictions of \citet{1984JGR....89.2673W} for non-resonant modes.
In our simulations, the speeds converge such that
the relative drift is roughly an order of magnitude smaller than its
initial value. \citet{2009ApJ...694..626R} suggest that convergence
to a relative velocity approximating the Alfv\'en speed is sufficient
to inhibit further magnetic-field amplification and thus that
saturation may take place. Although defining the Alfv\'en speed
for a turbulent magnetic field is complicated, we find that by
considering only the
projection of the magnetic field onto the drift direction along with
the local plasma density averaged over regions of a few $\lambda_{se}^2$
in area,
we do observe a ``mean'' Alfv\'en speed parallel to the cosmic-ray
drift that differs from the post-saturation velocity separation only
by a factor of order unity.

\begin{figure}[t]
  \centering
  \includegraphics[width=3in]{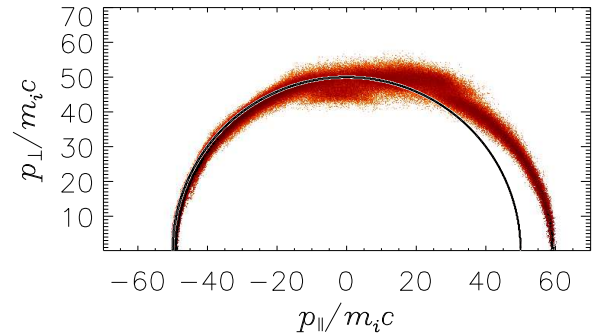}
  \caption{The phase-space distribution of cosmic rays in the instantaneous rest
frame of the population at time $t\gamma_{max}\approx 15.7$ in Run B, compared
with the initial distribution (thin semicircle). In addition to revealing the
anisotropy in the forms of stretching and broadening that began during the linear magnetic-field growth,
the instantaneous rest frame differs from the initial frame by approximately $0.18 c$.}
  \label{cr}
 \end{figure}

In addition to slowing down in bulk, the instantaneous rest-frame
distribution of cosmic rays is modified, as shown in Figure \ref{cr}.
Some anisotropy is introduced, which may have consequences for the
modeling both of radiation and particle acceleration. The anisotropy
appears in all simulations in a qualitatively similar manner, though
the rate at which it develops (as well as the extent to which
the CR bulk speed changes) decreases with increasing $\gamma_{CR}$.
The fastest development of  rest-frame anisotropy
occurs for cosmic rays whose trajectory is transverse to the drift
direction in the simulation. During the linear stage of magnetic-field
growth, the stationary parallel-wave mode with
wavelength $\lambda_{max}$ will
be favorable for resonant scattering of cosmic rays whose
pitch angle $\mu$ is given by $\lambda_{max} \Omega_{CR}/2\pi= v_{\parallel}=\mu\,c\left(1-\gamma^{-2}_{CR}\right)^{1/2}$.
For the parameters in our simulations  and in agreement
with the work of \citet{2009MNRAS.397.1402L}, $\mu$ is of order $10^{-2}$,
corresponding to those cosmic rays in which the
aforementioned anisotropy is first seen.
 Additional scattering and acceleration of
cosmic rays may occur in the electric fields
associated with magnetic fields being transported
in the turbulent flow during the nonlinear stage.
Second-order Fermi acceleration is possible in such
a turbulent environment. The electric fields we observe during
saturation are typically an order of magnitude smaller than the
magnetic fields. A Lorentz transformation into the $global$ plasma
rest frame reduces the electric fields only from about $t\gamma_{max}=15$
onward, in agreement with the behavior of the plasma bulk speed.
If we transform
the EM fields into the $local$ plasma rest frame, thereby accounting for its
turbulent motion, we see a reduction of the electric fields beginning at
$t\gamma_{max}\approx 10$. After about $t\gamma_{max}=15$, less than 20\% of the original electric field
remains, demonstrating that the dominant fraction of the electric field in
our simulation arises from the transport of magnetic fields.
This indicates that pure elastic scattering may be a poor assumption,
if one considers the diffusive transport of high-energy
cosmic rays in the upstream region of young SNR. Runs C--F with varying
cosmic-ray Lorentz factors illustrated principally that ``stiffer''
cosmic rays took slightly longer to begin slowing down and were
marginally more resistant to the onset of rest-frame anisotropy,
but there was no qualitative difference in the linear growth of
the magnetic field nor in the saturation mechanism.

The interactions between the shock
itself and any spatial variations not only in plasma
density but also in velocity are known to introduce vorticity
in the downstream region that may amplify the magnetic field
\citep{2007ApJ...663L..41G,2009ApJ...695..825I}.
They may also have a non-negligible impact on the
injection process or the scattering properties of
the downstream medium, and consequently
the resulting distribution of energetic particles \citep{2001RPPh...64..429M}.
If the injection
mechanism depends primarily on highly localized processes, as
opposed to being a product of the average conditions over large
spatiotemporal scales, then the presence of appropriately
distributed inhomogeneities might reduce the availability of readily
accelerated particles in one region, with no guarantee of a corresponding
increase in a neighboring region. In particular, if injection
is ultimately more efficient at a parallel shock than a perpendicular
shock, the development of a turbulent magnetic field in the
formerly parallel region via this streaming instability would
reduce the overall quasi-parallel fraction of the shock's surface
if comparable turbulence is not established in the initially
quasi-perpendicular regions.  Beyond injection, the cosmic-ray
transport properties upstream and downstream contribute
significantly to the acceleration efficiency and the maximum
energy cosmic rays can attain. \citet{2006A&A...453..193M} investigated the
sensitivity of the cosmic-ray diffusion coefficient to
the character of the turbulence, and found that, in particular,
anisotropic turbulence may lead to somewhat higher maximum particle
energy.

In addition to possible implications for the spectrum of
accelerated particles, the character of the upstream
magnetic turbulence also eventually contributes to the emission properties of
the downstream medium. The presence of anisotropy in the
turbulence after being shock-compressed may lead to a larger
degree of polarization in the radio synchrotron emission than
might otherwise be expected \citep{2009ApJ...696.1864S}.

The turbulent motion in the upstream medium may
be significant for self-consistent models of cosmic-ray
acceleration. In our simulations, plasma parcels attain velocities
that exceed the initial Alfv\'en speed and collide, producing
``fronts'' whose effects may include further localized field amplification
or, perhaps quite importantly, compression and
heating of the upstream medium. We observe that heating of
the ambient plasma occurs cospatially with the interfaces
between rapidly moving plasma parcels and the slower regions
into which they are pressing. In Run B, for instance, the average
kinetic energy per ion is of order $10^{-6}m_i c^2$ at
$t\gamma_{max}=3.8$, before the linear growth has begun,
and is slightly higher but still of that order at $t\gamma_{max}=7.6$,
during linear growth. But at $t\gamma_{max}=11.5$, the beginning of
saturation, the mean kinetic energy per ion has increased to
$\sim 5 \times 10^{-4} m_i c^2$, and at $t\gamma_{max}=15.3$,
it has climbed to $10^{-2} m_i c^2$.
The common and convenient assumption that the entire amplification and
saturation process
proceeds in a low-temperature regime must be carefully considered
if conditions allow for turbulence to develop rapidly
in the precursor,  but as \citet{2008ApJ...688.1084V} observe, the substantial
heating expected from the dissipation of even a
small fraction of upstream turbulence may significantly
increase the rate at which particles are injected into the
acceleration process.

The reduction of the cosmic-ray current (by the converging
drift velocities of their population and the ambient plasma) serves
to underscore the need to treat these particles kinetically. The
back-reaction from the turbulent plasma and amplified magnetic field
is non-negligible, with the most prominent consequence being
saturation of the field growth at a lower amplitude than might be expected
from a constant cosmic-ray current \citep{2009ApJ...698..445O}.

Finally, it is worth noting that although the initial dominant mode of
magnetic-field amplification agrees with the calculations of B04,
the later evolution and saturation mechanism have much in common with
the simulations performed in N08, in which the choice of simulation
parameters exposed an oblique filamentary mode as the dominant instability.
It seems it doesn't matter what form the
initial linear instability takes.

{\ts An important consideration in interpreting our findings is} that the ability
of our simulations to accurately portray the evolution of the magnetic field
and
the plasma properties at very late times, beyond the saturation,
is limited. In addition to the
limitation imposed by the finite size of the simulation domain,
the assumptions of environmental homogeneity become increasingly unsound.
Changes in the plasma flow speed imply density adjustments in order
to conserve mass flux \citep{2009ApJ...698..445O}. Also, as the shock approaches, spatial
gradients in the cosmic-ray precursor and ambient plasma properties
play an increasingly important role. Furthermore, the time in which
upstream amplification can occur before the system is overtaken
by the shock is only $\sim 40\gamma_{max}^{-1}$ for our parameters,
assuming efficient Bohm-type diffusion so that the shock sweep-up
timescale is $c\,r_{CRg}/3v_{sh}^2$
(where $r_{CRg}$ is the cosmic-ray gyroradius) (N08).
Thus, field growth is inevitably limited even in the absence of saturation.

{\ts It bears mentioning that to a first approximation, the
effects of a changing environment will be limited. As mentioned
in \S \ref{model}, the parameters of the streaming instability are agnostic
to the details of the cosmic-ray distribution. Thus, although an approaching
precursor implies a continual softening of the local cosmic-ray energy
spectrum due to the arrival of particles over an increasingly
inclusive energy range, it is only the increased overall flux that matters.
Until surplus high-energy electrons are included, the net cosmic-ray
current will increase, whose effect according to Equation \ref{instab}
will be to amplify the parallel mode described therein on shorter time- and
length-scales, potentially well separated from the large scales
attained by the turbulence generated earlier.  Even this effect
is difficult to predict with confidence: not only will cosmic rays that
arrive during saturation encounter a turbulent upstream medium, but may
themselves have been influenced by prior unstable evolution nearer
to the shock. In any case, since the precursor length scale is proportional to
the diffusion coefficient, and hence to the energy for Bohm diffusion, cosmic
ray particles of significantly lower energy (and higher flux) are brought in
only shortly prior to capture by the shock.

Although the non-resonant streaming instability is concerned
only with the current they carry (and, as demonstrated by
\citet{2006A&A...453..181P} and \citet{2009MNRAS.397.1402L}, is largely
produced by the return current in the plasma), this or other instabilities
thus encountered may introduce anisotropy and inhomogeneity into the
local cosmic-ray distribution, with potentially non-trivial consequences. As increased
computational capacity and improved models become available, we can
expect simulations
representing a significant extent of the precursor to yield valuable
insight into these interrelated and nonlinear processes occurring there.}

\section{Summary and conclusions}
\label{conclusions}
We have simulated the turbulent amplification of the interstellar magnetic field
upstream of a nonrelativistic shock
using a kinetic 2.5-dimensional particle-in-cell code.

We observe:
\begin{itemize}
\item That the non-resonant streaming instability (B04)
is seen initially
for our choice of parameters, but with fluctuations appearing in
the plasma density;

\item That the plasma quickly evolves and saps
the bulk momentum from the drifting cosmic rays;

\item The reduction of relative drift effectively removes
the cosmic-ray current and saturates the magnetic-field amplification to
$\sim 20 B_{\parallel 0}$;

\item This saturation mechanism is independent
of the initial linear instability,
occurring as a result of the back-reaction on the cosmic rays;

\item Strong transverse plasma motions arise in
conjunction with the turbulent magnetic
field exceeding the homogeneous background field;

\item Interactions between differently-moving plasma parcels
give rise to significant fluctuations in density and temperature
of the upstream interstellar medium;

\item The plasma begins to drift,
suggestive of a cosmic ray-modified shock, until its speed approximately matches
that of the cosmic-ray population.

\item The phase-space distribution of cosmic rays undergoes significant
changes: the development of anisotropy within the population rest frame
beginning during the linear growth of the magnetic field, and decrease in bulk
speed associated temporally with the saturation of the magnetic field;

\item Evolution past this point is beyond
the reach of our current simulation method, as it requires additional input
to reflect the changing external environment.
\end{itemize}
The growth of non-resonant streaming instabilities in the cosmic-ray
precursor of supernova shocks may contribute to significant turbulent
amplification of the magnetic field upstream of the shock. However,
the amplitudes achieved via any one process are likely to be
somewhat less than suggested in the artificial case of
constant cosmic-ray current. Simulations that incorporate incrementally
more of the environmental changes expected in a real precursor
scenario can be expected as computational capacity increases,
although it may take innovations beyond the foreseeable future
to make a realistic particle-in-cell approach tractable for such
a simulation \citep{2008ApJ...688.1084V}.

This research was supported in part by the National Science Foundation
 both through TeraGrid resources provided by NCSA \citep{Teragrid}  and
under Grant No. PHY05-51164.
The work of JN is supported
by MNiSW research project N N203 393034, and The Foundation for Polish
Science through the HOMING program, which is supported by a grant from
Iceland, Liechtenstein, and Norway through the EEA Financial
Mechanism.

\bibliographystyle{astron}
\bibliography{streaming}

\begin{thebibliography}{}

\bibitem[\protect\astroncite{{Bell}}{2004}]{2004MNRAS.353..550B}
{Bell}, A.~R.: 2004,
\newblock {\em \mnras} {\bf 353}, 550

\bibitem[\protect\astroncite{{Buneman}}{1993}]{1993cspp.book......M}
{Buneman}, O.: 1993,
\newblock {\em {Computer Space Plasma Physics: Simulation Techniques and
  Software, ed. H. Matsumoto \& Y. Omura}}, pp 67--84,
\newblock ~Tokyo: Terra

\bibitem[\protect\astroncite{{Catlett,~C.~et~al.}}{2007}]{Teragrid}
{Catlett,~C.~et~al.}: 2007,
\newblock {\em {TeraGrid: Analysis of Organization, System Architecture, and
  Middleware Enabling New Types of Applications}},
\newblock IOS Press

\bibitem[\protect\astroncite{{Giacalone} and
  {Jokipii}}{2007}]{2007ApJ...663L..41G}
{Giacalone}, J. and {Jokipii}, J.~R.: 2007,
\newblock {\em \apjl} {\bf 663}, L41

\bibitem[\protect\astroncite{{Greenwood} et~al.}{2004}]{2004JCoPh.201..665G}
{Greenwood}, A.~D., {Cartwright}, K.~L., {Luginsland}, J.~W., and {Baca},
  E.~A.: 2004,
\newblock {\em Journal of Computational Physics} {\bf 201}, 665

\bibitem[\protect\astroncite{{Inoue} et~al.}{2009}]{2009ApJ...695..825I}
{Inoue}, T., {Yamazaki}, R., and {Inutsuka}, S.-i.: 2009,
\newblock {\em \apj} {\bf 695}, 825

\bibitem[\protect\astroncite{{Luo} and {Melrose}}{2009}]{2009MNRAS.397.1402L}
{Luo}, Q. and {Melrose}, D.: 2009,
\newblock {\em \mnras} {\bf 397}, 1402

\bibitem[\protect\astroncite{{Malkov} and {Drury}}{2001}]{2001RPPh...64..429M}
{Malkov}, M.~A. and {Drury}, L.~O.: 2001,
\newblock {\em Reports on Progress in Physics} {\bf 64}, 429

\bibitem[\protect\astroncite{{Marcowith} et~al.}{2006}]{2006A&A...453..193M}
{Marcowith}, A., {Lemoine}, M., and {Pelletier}, G.: 2006,
\newblock {\em \aap} {\bf 453}, 193

\bibitem[\protect\astroncite{{Niemiec} et~al.}{2008}]{2008ApJ...684.1174N}
{Niemiec}, J., {Pohl}, M., {Stroman}, T., and {Nishikawa}, K.-I.: 2008,
\newblock {\em \apj} {\bf 684}, 1174

\bibitem[\protect\astroncite{{Ohira} et~al.}{2009}]{2009ApJ...698..445O}
{Ohira}, Y., {Reville}, B., {Kirk}, J.~G., and {Takahara}, F.: 2009,
\newblock {\em \apj} {\bf 698}, 445

\bibitem[\protect\astroncite{{Pelletier} et~al.}{2006}]{2006A&A...453..181P}
{Pelletier}, G., {Lemoine}, M., and {Marcowith}, A.: 2006,
\newblock {\em \aap} {\bf 453}, 181

\bibitem[\protect\astroncite{{Reville} et~al.}{2009}]{2009ApJ...694..951R}
{Reville}, B., {Kirk}, J.~G., and {Duffy}, P.: 2009,
\newblock {\em \apj} {\bf 694}, 951

\bibitem[\protect\astroncite{{Reynolds}}{2008}]{2008ARA&A..46...89R}
{Reynolds}, S.~P.: 2008,
\newblock {\em \araa} {\bf 46}, 89

\bibitem[\protect\astroncite{{Riquelme} and
  {Spitkovsky}}{2009}]{2009ApJ...694..626R}
{Riquelme}, M.~A. and {Spitkovsky}, A.: 2009,
\newblock {\em \apj} {\bf 694}, 626

\bibitem[\protect\astroncite{{Stroman} and {Pohl}}{2009}]{2009ApJ...696.1864S}
{Stroman}, W. and {Pohl}, M.: 2009,
\newblock {\em \apj} {\bf 696}, 1864

\bibitem[\protect\astroncite{{Umeda} et~al.}{2003}]{2003CoPhC.156...73U}
{Umeda}, T., {Omura}, Y., {Tominaga}, T., and {Matsumoto}, H.: 2003,
\newblock {\em Computer Physics Communications} {\bf 156}, 73

\bibitem[\protect\astroncite{{Vladimirov} et~al.}{2008}]{2008ApJ...688.1084V}
{Vladimirov}, A.~E., {Bykov}, A.~M., and {Ellison}, D.~C.: 2008,
\newblock {\em \apj} {\bf 688}, 1084

\bibitem[\protect\astroncite{{Winske} and {Leroy}}{1984}]{1984JGR....89.2673W}
{Winske}, D. and {Leroy}, M.~M.: 1984,
\newblock {\em \jgr} {\bf 89}, 2673

\bibitem[\protect\astroncite{{Zirakashvili} et~al.}{2008}]{2008ApJ...678..255Z}
{Zirakashvili}, V.~N., {Ptuskin}, V.~S., and {V{\"o}lk}, H.~J.: 2008,
\newblock {\em \apj} {\bf 678}, 255

\end{thebibliography}

\end{document}